\documentclass[showpacs,preprintnumbers,amsmath,amssymb,APSl,prd,nofootinbib,superscriptaddress, twocolumn]{revtex4}

\usepackage{dcolumn}% Align table columns on decimal point
\usepackage{bm}% bold math
\usepackage{ifpdf}
\usepackage{hyperref}
\usepackage{bm}
\usepackage{xcolor,color,graphicx,graphics}
\usepackage[spanish,english]{babel}%, portuguese
\usepackage[latin1]{inputenc}
\usepackage[OT1]{fontenc}
\usepackage{latexsym,amssymb,amsmath,amsfonts}
\usepackage{makeidx}
\usepackage{epsfig,subfigure}
\usepackage{natbib}
\usepackage{epstopdf}
\usepackage{mathrsfs}
\hypersetup{colorlinks=true, linkcolor=blue, citecolor=green}
\usepackage{enumerate}
\usepackage{xcolor}
 \usepackage{multirow}
 
\definecolor{red}{rgb}{1,0,0}

\def\+{^\dagger}

\def\<{\leftarrow}
\def\>{\rightarrow}

\def\({\left(}
\def\){\right)}

\newcommand{\bi}{\begin{itemize}} 				\newcommand{\ei}{\end{itemize}}
\newcommand{\benu}{\begin{enumerate}} 		\newcommand{\enu}{\end{enumerate}}
\newcommand{\bd}{\begin{dinglist}{0}}     \newcommand{\ed}{\end{dinglist}}
\newcommand{\bfig}{\begin{figure}[htbp]}  \newcommand{\efig}{\end{figure}}
        			
\newcommand{\bc}{\begin{center}} 				  \newcommand{\ec}{\end{center}}
\newcommand{\be}{\begin{equation}} 				\newcommand{\ee}{\end{equation}}
\newcommand{\bsub}{\begin{subequations}}  \newcommand{\esub}{\end{subequations}}
\newcommand{\ba}[1]{\begin{array}{#1}} 		\newcommand{\ea}{\end{array}}
\newcommand{\bea}{\begin{eqnarray}}
\newcommand{\eea}{\end{eqnarray}}

\newcommand{\nn}{\nonumber \\}
\begin{document}
\title{Boundary terms and on-shell action in Ricci-based gravity theories: the Hamiltonian formulation}

	\author{Gerardo Mora-P\'erez}
\affiliation{Departamento de F\'{i}sica Te\'{o}rica and IFIC, Centro Mixto Universidad de Valencia - CSIC.
Universidad de Valencia, Burjassot-46100, Valencia, Spain}
\author{Gonzalo J. Olmo} \email{gonzalo.olmo@uv.es}
\affiliation{Departamento de F\'{i}sica Te\'{o}rica and IFIC, Centro Mixto Universidad de Valencia - CSIC.
Universidad de Valencia, Burjassot-46100, Valencia, Spain}
\affiliation{Departamento de F\'isica, Universidade Federal do Cear\'a (UFC), Campus do Pici, Fortaleza - CE, C.P. 6030, 60455-760 - Brazil}
\author{Diego Rubiera-Garcia} \email{drubiera@ucm.es}
\affiliation{Departamento de F\'isica Te\'orica and IPARCOS,
	Universidad Complutense de Madrid, E-28040 Madrid, Spain}
	\author{Diego S\'aez-Chill\'on G\'omez} \email{diego.saez@uva.es}
\affiliation{Department of Theoretical Physics, Atomic and Optics, Campus Miguel Delibes, \\
University of Valladolid UVA, Paseo Bel\'en, 7, 47011 - Valladolid, Spain}
\affiliation{Departamento de F\'isica, Universidade Federal do Cear\'a (UFC), Campus do Pici, Fortaleza - CE, C.P. 6030, 60455-760 - Brazil}

\date{\today}
\begin{abstract}
Considering the so-called Ricci-based gravity theories, a family of extensions of General Relativity whose action is given by a non-linear function of contractions and products of the (symmetric part of the) Ricci tensor of an independent connection, the Hamiltonian formulation of the theory is obtained. To do so, the independent connection is decomposed in two parts, one compatible with a metric tensor and the other one given by a  3-rank tensor. Subsequently, the Riemann tensor is expressed in terms of its projected components onto a hypersurface, allowing to construct the $3+1$ decomposition of the theory and the corresponding Gauss-Codazzi relations, where the boundary terms naturally arise in the gravitational action. Finally, the ADM decomposition is followed in order to construct the corresponding Hamiltonian and the ADM energy for any Ricci-based gravity theory. The formalism is applied to the simple case of Schwarzschild space-time.

\end{abstract}

\maketitle

%%%%%%%%%%%%%%%%%%%%%%%%%%%%%%%%%%%%%%%%
\section{Introduction}
%%%%%%%%%%%%%%%%%%%%%%%%%%%%%%%%%%%%%%%%

Over the last  hundred years, General Relativity (GR) has shown its incredible power of predictability by satisfying all observational constraints \cite{Will:2014kxa} and predicting new phenomena that have been later observed, as the gravitational waves of binary mergers \cite{LIGOScientific:2016aoc} and the shadows of supermassive black holes \cite{EventHorizonTelescope:2019dse,EventHorizonTelescope:2022wkp}. The theory describes gravity as a geometric phenomenon, based on the equivalence principle and the principle of general covariance, such that a manifold that represents the physical space-time is defined by a metric and a connection that provide the length of vectors and the rule for parallel transport, respectively. However, while the central object in the original and standard formulation of GR is the curvature of the metric, which assumes the Levi-Civita connection compatible with the metric, the theory can be formulated in (at least)  two other  completely equivalent ways: firstly, by assuming the presence exclusively of torsion (Teleparallel gravity) and secondly, by assuming the presence exclusively of non-metricity (coincident General Relativity). This is popularly known as the trinity of gravity to formulating GR \cite{BeltranJimenez:2019esp}. 

For every formulation of gravity above, the connection is fixed a priori. Nevertheless, no fundamental principle requires a particular choice of the connection but it can be instead taken as an additional free field. This is the basis of the so-called Palatini formalism, which for the action of GR yields the Levi-Civita connection from resolving the corresponding field equations for the connection, thus being equivalent to the canonical metric-based formulation. However, other gravitational actions different from GR and involving the (symmetric part of the) Ricci tensor lead to a non-compatible connection with the metric. These are generally known as Ricci Based Gravity (RBG) theories \cite{Afonso:2018bpv}, and are a subset of the more general family of metric-affine theories of gravity.

%Originally inspired by non-linear Born-Infeld electrodynamics applied in the gravitational sector, 
RBGs have shown their potential as the departing point to construct a more suitable theory of gravity \cite{BeltranJimenez:2017doy}. Indeed, it has been generally argued that GR is not a complete theory given the fact that it harbors space-time singularities in some of its most physically appealing solutions (such as black holes or the early cosmological evolution), and furthermore it does not naturally incorporate quantum mechanics on its strong-field regime. However, when the Einstein-Hilbert action of GR is generalized to incorporate higher-order curvature terms, it usually leads to fourth-order field equations and the presence of instabilities (ghosts). Nevertheless, those modifications of the GR action formulated within the Palatini formalism keep the second-order of the field equations, avoiding in this way the presence of ghosts (for reviews see \cite{BeltranJimenez:2017doy,Olmo:2011uz,Olmo:2020fnk}). The simplest version of these theories is the $f(R)$ case, which just considers a general function of the scalar $R=g^{\mu\nu}R_{\mu\nu}(\Gamma)$ in the action, where the Ricci tensor depends on the independent connection $\Gamma$. Such theories have been widely analyzed, mainly within the cosmological framework in hopes that they might provide an explanation to dark energy, inflation, and other phenomena \cite{Baghram:2009we,Aoki:2018lwx,Leanizbarrutia:2017xyd,Rosa:2017jld,Rosa:2021ish,Harko:2011nh,Shimada:2018lnm,Gialamas:2020snr,Antoniadis:2018ywb,Rasanen:2018ihz,Bekov:2020dww,Karam:2021sno,Koivisto:2005yc}, but intensive work has been also done within the context of compact objects and, particularly, on regular black holes \cite{Olmo:2015axa,Guerrero:2020uhn,Rosa:2020uoi}. More formal aspects of these theories have been also studied, see e.g. \cite{Odintsov:2014yaa,Olmo:2019flu,Olmo:2019flu,Capozziello:2010ut,BeltranJimenez:2018vdo,Martinez-Asencio:2012ebc}. As a final token of these theories, it turns out that RBGs can be mapped into a GR-like form, such that all the possible solutions and properties of GR can be directly extrapolated to RBG theories but where new features for the space-time solutions might arise due to the new gravitationally-induced interactions \cite{Afonso:2018bpv}. Note also that such mapping technique allows one to analyze complex aspects and frameworks for these theories \cite{Afonso:2019fzv,Delhom:2019zrb,Afonso:2018hyj} and also to carry out demanding numerical relativity calculations. 

There is one issue that has remained unsolved over all these years though, namely, the development of a well-defined Hamiltonian formulation for any RBG. The main pitfall behind lied on formal aspects, and mainly on establishing a well-defined variational principle for these theories. The problem has been extensively analysed in GR, where also some issues arise when considering null hypersurfaces \cite{Parattu:2015gga}, and new techniques have been developed over the last years \cite{Chakraborty:2016yna}. The point in RBG theories is that as far as one assumes fixed values of the fields at the boundaries of the volume over which the action is integrated, the boundaries terms, as the Gibbons-Hawking-York (GHY) term of GR, are removed, since variations of the connection turn out null at the boundaries. Nevertheless, the presence of such boundary terms have been shown for the case of Palatini $f(R)$ gravities to be fundamental, among other stuff, to recover the correct expression for the Schwarzschild black hole entropy \cite{Saez-ChillonGomez:2020afj}. In the same way, such boundary terms, analogs of the GHY term, have been obtained by different approaches \cite{BarberoG:2021cei,Gomez:2021roj}.

The aim of this paper is to go one step further in the enlightening of this topic with respect to previous analyses. Indeed, here we shall consider a completely general RBG theory and we settle to obtain first the corresponding boundary terms in the gravitational action and, second, to establish a well-defined Hamiltonian formalism for this class of theories. To do so, the mapping of RGB into GR-like form is used in order to transform the general gravitational action into one with just linear terms of the Ricci tensor \cite{Afonso:2018bpv}. Then, by decomposing the free symmetric connection into two parts, a fiducial connection compatible with a metric plus a rank-3 tensor, the Riemann and Ricci tensors are expressed in terms of these two objects for foliating the corresponding manifold into a family of hypersurfaces. The latter allows to express such tensors in terms of extrinsic and intrinsic magnitudes of the hypersurfaces, the so-called Gauss-Codazzi equations (also known as the $3+1$ decomposition), already implemented successfully in Palatini $f(R)$ gravities \cite{Gomez:2021roj}. Finally, the Hamiltonian formulation is obtained by following the standard procedure of implementing the Arnowitt-Deser-Misner (ADM) decomposition. Results mimic GR formulation, as expected due to the mapping. However, important different features arise, as shown when computing the ADM energy for the Schwarzschild black hole.  Moreover, one should note that the same boundary terms can be obtained by applying variations directly into the action without any mapping to GR. For the latter, the crucial point is to decompose the connection into two parts, one representing a fiducial connection compatible with an unspecified metric tensor, and a second part that consists on a rank-3 tensor whose field equations lead to a trivial result.

The paper is organized as follows: in Sec.  \ref{RGBtheories} the RGB theories are introduced. Sec. \ref{31decomposition} is devoted to obtain the $3+1$ decomposition and the corresponding Gauss-Codazzi action that leads to the analog of the GHY boundary term. In sec. \ref{CovariantSect}, the same boundary terms are also obtained by following a different approach. Sec. \ref{Hamiltonian} establishes the general Hamiltonian formulation for RGB theories and the ADM energy is calculated. Finally, Sec. \ref{conclusions} gathers the conclusions of the paper.

%%%%%%%%%%%%%%%%%%%%%%%%%%%%%%%%%%%%%%
\section{Variational principle in Ricci-based theories of gravity}
\label{RGBtheories}
%%%%%%%%%%%%%%%%%%%%%%%%%%%%%%%%%%%%%%

Let us consider a theory of gravity defined as
\begin{eqnarray} 
\mathcal{S}&=&\mathcal{S}_G + \mathcal{S}_M=\frac{1}{2\kappa^2} \int d^4x \sqrt{-g} F(g_{\mu\nu},R_{\mu\nu}(\Gamma)) \nonumber \\
& + & \int d^4x \sqrt{-g} \mathcal{L}_m (g_{\mu\nu},\psi_m)\ , \label{eq:action}
\end{eqnarray}
where $F$ is a given scalar built out from the metric and the symmetric part of the Ricci tensor. The latter requirement is needed in order to avoid troubles with ghost-like instabilities \cite{BeltranJimenez:2019acz}.
The Ricci-dependent part of this action can be linearized by considering the alternative representation \cite{BeltranJimenez:2017doy,Rubiera-Garcia:2020gcl} 
\begin{eqnarray} 
\mathcal{S}_G&=& \int d^4x \frac{\sqrt{-g}}{2\kappa^2} \left(\frac{\partial F}{\partial \Sigma_{\mu\nu}}R_{\mu\nu}(\Gamma)+F(g,\Sigma)- \Sigma_{\mu\nu}\frac{\partial F}{\partial \Sigma_{\mu\nu}}\right) , \label{eq:actionSigma}
\end{eqnarray}
which involves the non-dynamical field $\Sigma_{\mu\nu}$. One can check that the variation over $\Sigma_{\mu\nu}$ implies that $\Sigma_{\mu\nu}=R_{\mu\nu}$, making the above action dynamically equivalent to $\mathcal{S}_G$ in (\ref{eq:action}) .
% For our purposes, we shall also set torsion to zero in order to simplify the analysis, though it can be simply accommodated into the general formalism \cite{Afonso:2017bxr}.
Using this representation, one can now define an auxiliary metric $q_{\mu\nu}$ via the following relation (for details on this procedure see e.g. \cite{BeltranJimenez:2017doy,Rubiera-Garcia:2020gcl}):
\begin{equation}
\sqrt{-q}q^{\mu\nu} \equiv \sqrt{-g}\frac{\partial F}{\partial \Sigma_{\mu\nu}}\ ,
\label{metricrselation}
\end{equation}
so that the action (\ref{eq:action}) reads
\begin{equation} \label{eq:actionRBG}
\mathcal{S}_{RBG} = \frac{1}{2\kappa^2} \int d^4x \sqrt{-q} q^{\mu\nu}R_{\mu\nu}(\Gamma) + \tilde{\mathcal{S}}_m (g_{\mu\nu},\psi_m)\ .
\end{equation}
The action expressed in this way looks like the Palatini version of GR for the auxiliary metric $q_{\mu\nu}$  but with a modified matter sector $\tilde{\mathcal{S}}_m$ that results from combining the original $ \mathcal{S}_M$ and the extra terms $F(g,\Sigma)- \Sigma_{\mu\nu}\frac{\partial F}{\partial \Sigma_{\mu\nu}}$ that appear in  (\ref{eq:actionSigma}). Assuming that the dependence on the original metric $g_{\mu\nu}$ can be written in terms of $q_{\mu\nu}$ and the other fields (see for instance \cite{Afonso:2018bpv}), the field equations for the metric $q_{\mu\nu}$ and the connection can be obtained by variation of the action (\ref{eq:actionRBG}) with respect to them as
\begin{eqnarray}
\delta \mathcal{S} &=&\frac{1}{2\kappa^2}\int d^4x \left[ \frac{\delta\left(\sqrt{-q}q^{\mu\nu}\right)}{\delta q^{\alpha\beta}}R_{\mu\nu} \delta q^{\alpha\beta}  + \sqrt{-q}q^{\mu\nu} \delta R_{\mu\nu} \right] \nonumber \\
&+& \delta \tilde{\mathcal{S}}_m(q_{\mu\nu},\psi_m) \label{eq:var}\ .
\end{eqnarray}
There are two pieces clearly separated in this variation because metric and connection are independent; those related to the metric yield the corresponding field equations, while those related to the Ricci tensor yield the ones associated to the connection. Indeed, the former piece leads to the field equations for the metric $q$ as
\be
R_{\mu\nu}(\Gamma)-\frac{1}{2}q_{\mu\nu}R(\Gamma)=\kappa^2 \tilde{T}_{\mu\nu}\ ,
\label{eqsMetricq}
\ee
where $R(\Gamma)=q^{\alpha\beta}R_{\alpha\beta}(\Gamma)$ is the curvature scalar of the independent connection while $\tilde{T}_{\mu\nu}=\frac{-2}{\sqrt{-q}}\frac{\delta\tilde{\mathcal{S}}_m}{\delta q^{\mu\nu}}$ is the (effective) stress-energy tensor of the matter fields.  %As stated above, these are Einstein equations for the metric $q_{\mu\nu}$ and a modified energy-momentum tensor  $\tilde{T}_{\mu\nu}$ (in this formulation of the action).

The latter piece in the variation (\ref{eq:var}) is the one we are most interested in. Indeed, the variation with respect to the connection can be separated into two contributions: the ones leading to the field equations for the connection and the associated boundary contributions. To work out both contributions, we call upon the expression of the variation of the Ricci tensor (in this torsionless case\footnote{We note here that the result including torsion does not change anything due to the projective invariance of the theory. The reader is invited to read \cite{Afonso:2017bxr} and \cite{Olmo:2011uz} for a detailed derivation in that case.}) as
\begin{equation}
\delta R_{\mu\nu}= \nabla_{\lambda} \delta \Gamma_{\mu\nu}^{\lambda} - \nabla_{\nu} \delta \Gamma_{\lambda \mu}^{\lambda}\ .
\end{equation}
By using this expression, the piece on the variation of the connection in (\ref{eq:var}) can be expressed as follows:
\begin{eqnarray}
\delta \mathcal{S}_{\Gamma}(\Gamma) &=&\frac{1}{2\kappa^2 } \int d^4x \sqrt{-q}q^{\mu\nu} \delta R_{\mu\nu} \nonumber \\
&=&   \frac{1}{2\kappa^2} \int d^4x \sqrt{-q}q^{\mu\nu} \left[ \nabla_{\lambda} \delta \Gamma_{\mu\nu}^{\lambda} - \nabla_{\nu} \delta \Gamma_{\lambda \mu}^{\lambda} \right]\ .
\end{eqnarray}
Then, the variation can be integrated by parts and split the result as $\delta \mathcal{S}_{\Gamma}(\Gamma)=\delta \mathcal{S}_{\Gamma}^I+\delta \mathcal{S}_{\Gamma}^{II}$ where each piece reads as
\begin{equation}  \label{eq:conn1}
\delta \mathcal{S}_{\Gamma}^I  =  \int d^4x  \frac{\sqrt{-g} }{2\kappa^2}\nabla_{\lambda}  \Big( \sqrt{-q} q^{\mu\nu} \delta \Gamma_{\mu\nu}^{\lambda} -\sqrt{-q}q^{\mu\lambda} \delta \Gamma_{\rho \mu}^{\rho} \Big)
\end{equation}
and
\begin{equation}
\delta \mathcal{S}_{\Gamma}^{II} = -  \int d^4x  \frac{\sqrt{-g} }{2\kappa^2} \delta \Gamma_{\mu\nu}^{\lambda} \Big(\nabla_{\lambda} [\sqrt{-q}q^{\mu\nu} ] - \delta_{\lambda}^{\nu} \nabla_{\rho} [\sqrt{-q} q^{\mu\rho} ] \Big)\ \label{eq:conn2}
\end{equation}
We thus see that Eq. (\ref{eq:conn1}) represents the boundary terms of the gravitational action, while  (\ref{eq:conn2}) provides the field equations for the connection, which can be written as
\begin{equation}
\nabla_{\lambda} [\sqrt{-q}q^{\mu\nu} ] - \delta_{\lambda}^{\nu} \nabla_{\rho} [\sqrt{-q} q^{\mu\rho} ] = \frac{\delta \tilde{\mathcal{S}}_m}{\delta \Gamma_{\mu\nu}^{\lambda}}\ ,
\end{equation}
On the right-hand side of this expression we have included a term that would arise if we had considered connection-dependent terms on the matter sector of the action (\ref{eq:action}) (e.g. fermionic fields or non-minimal couplings). Since, for simplicity, such terms are absent in the analysis of this work, we set it to zero, $ \frac{\delta \tilde{\mathcal{S}}_m}{\delta \Gamma_{\mu\nu}^{\lambda}}=0$. Next, by contracting the indices $\lambda$ and $\nu$ one finds that the second term on the left-hand side vanishes, leading to \begin{equation} \label{eq:connsol}
\nabla_{\lambda} [\sqrt{-q} q^{\mu\nu} ]=0\ ,
\end{equation}
which tells us that the independent connection $\Gamma$ is actually the Levi-Civita connection of $q_{\mu\nu}$, i.e., it is given by the Christoffel symbols of the auxiliary metric. This is a well-known result of RBGs, which we recover here as a side effect of our treatment.

Let us now handle the boundary term (\ref{eq:conn1}). Since the object $\sqrt{-q}$ is a scalar density of weight $-1$, then its covariant derivative reads as
\begin{equation}
\nabla_{\mu} \sqrt{-q} =\partial_{\mu} \sqrt{-q} -\sqrt{-q} \Gamma_{\mu\sigma}^{\sigma}\ . 
\end{equation}
In absence of torsion, for any vector $V^{\sigma}$ one trivially has that $\nabla_{\sigma} (\sqrt{-q} V^{\sigma})=\partial_{\sigma}(\sqrt{-q} V^{\sigma})$. In Eq. (\ref{eq:conn1}) we identify this vector by:
\begin{equation}
V^{\sigma}= q^{\mu\nu}\delta \Gamma_{\mu\nu}^{\sigma} -q^{\mu\sigma} \delta \Gamma_{\lambda \mu}^{\lambda}\ .
\end{equation}
Applying Gauss' theorem in a certain volume $\mathcal{V}$, the integral of a divergence leads to
\begin{equation}\label{eq:WrongST}
\int_{\mathcal{V}} d^4x \sqrt{-q}\nabla_{\sigma} V^{\sigma}=\epsilon \int_{\delta\mathcal{V}} d^3x \sqrt{|h|} n_{\sigma}V^{\sigma}\ ,
\end{equation}
where $n^{\sigma}$ is the normal vector to the hypersurface $\delta\mathcal{V}$ enclosing such a volume, the constant $\epsilon= n_{\mu}n^{\mu}=\pm 1, 0$  depending on whether the hypersurface is time-like, space-like or null, respectively, and $h$ is the determinant of the induced $h$-metric on the boundary $\delta\mathcal{V}$. 

At this point, the variational approach followed above becomes useless for obtaining the corresponding boundary terms because the connection coefficients are fixed at the boundary and, consequently, their variations on the hypersurface $\delta\mathcal{V}$ vanish by definition, $\delta \Gamma=0$. Nevertheless, such boundary terms are present, as shown for instance in Ref.~\cite{Saez-ChillonGomez:2020afj,BarberoG:2021cei}. A possible way out of this seemingly contradiction is by considering the independent connection as being the Levi-Civita connection of an undetermined metric tensor, which becomes $q_{\mu\nu}$ after solving the field equations for the connection (\ref{eq:connsol}). However, such an approach would be against the philosophy of the variational principle, because it would be restricting the possible variations of the connection to those that satisfy the field equations. In fact, the variations of the connection can be represented by a rank-3 tensorial component whose variations cannot be completely encoded in the variations of a rank-2 tensor. For this reason, a more general approach is necessary to make progress in the understanding of this problem. Next we will proceed by decomposing the Ricci tensor in the $3+1$ formulation. An additional covariant discussion will be provided afterwards. 

%%%%%%%%%%%%%%%%%%%%%%
\section{$3+1$ decomposition and boundary terms}
\label{31decomposition}
%%%%%%%%%%%%%%%%%%%%%%
In order to establish the decomposition of the Ricci tensor and curvature in terms of the properties of a hypersurface (or a family of hypersurfaces), firstly we shall decompose the affine connection as
\begin{equation}\label{eq:GammaSplit}
\Gamma_{\mu\nu}^{\lambda}=L_{\mu\nu}^{\lambda} + A_{\mu\nu}^{\lambda}
\end{equation}
where $L_{\mu\nu}^{\lambda}$ represents a fiducial Levi-Civita connection built with the metric $q_{\mu\nu}$, while the rank-3 tensor $A_{\mu\nu}^{\lambda}$ contains all (so far undefined) extra terms. The choice of $q_{\mu\nu}$  as the metric associated to $L_{\mu\nu}^{\lambda}$ (instead of $g_{\mu\nu}$ or any other metric tensor) is  motivated by the resemblance that our effective action (\ref{eq:actionRBG}) has with respect to GR when formulated in terms of it and, obviously, from the fact that the field equations will choose such a connection, as given by Eq.(\ref{eq:connsol}). %However, additional variations may appear by the differences between $L_{\mu\nu}^{\lambda}$ and  $\Gamma_{\mu\nu}^{\lambda}$ due to the boundary terms, which are the ones we want to obtain here. 
Note that the decomposition (\ref{eq:GammaSplit}) puts forward that when a fiducial connection is defined, such as $L_{\mu\nu}^{\lambda}$ here, there exists a bijection between the difference of two connections and the space of rank-3 tensors. The space of connections is an affine space, while the space of tensors is a vector space and, generally, it is easier and clearer to work with vector spaces. That is why we have followed this approach and, as we will see, it will proof useful in this discussion. Under this decomposition of the affine connection, the Riemann tensor can be expressed as:
\bea
R^{\lambda}_{\;\;\sigma\mu\nu}(\Gamma)&=R^{\lambda}_{\;\;\sigma\mu\nu}(L)+\mathcal{A}^{\lambda}_{\;\;\sigma\mu\nu}\ ,
\label{RiemannTensor}
\eea
where $R^{\lambda}_{\;\;\sigma\mu\nu}(L)$ is the usual Riemann tensor in terms of the connection $L^{\lambda}_{\mu\nu}$, while the second term in (\ref{RiemannTensor}) is given by:
\be
\mathcal{A}^{\lambda}_{\;\;\sigma\mu\nu}=\nabla_{\mu}^L A_{\;\;\sigma\nu}^{\lambda} -\nabla_{\nu}^{L}A_{\;\;\sigma\mu}^{\lambda} +A_{\;\;\mu\rho}^{\lambda} A_{\;\;\nu\sigma}^{\rho}-A_{\;\;\nu\rho}^{\lambda}A_{\;\;\mu\sigma}^{\rho}\ .
\label{Atensor}
\ee
Here the superindex indicates that covariant derivatives are taken with respect to the Levi-Civita connection of $q$. Let us now obtain the Ricci tensor expressed in terms of the magnitudes defined for a family of hypersurfaces. Let us consider a foliation of the manifold defined by the metric tensor $q_{\mu\nu}$ into a (or a family of) hypersurface(s) $\Sigma$ defined by the normal vector
\be 
n_\mu=\frac{\partial_{\mu} S(x)}{\sqrt{|q^{\sigma\lambda}\partial_{\sigma} S(x)\partial_{\lambda} S(x)}|}\ , \quad S(x)=0\ .
\ee
Then, the induced metric tensor $h_{\mu\nu}$ is given by:
\begin{equation} \label{eq:inducedmetric1}
q_{\mu\nu}=h_{\mu\nu} + \varepsilon n_{\mu}n_{\nu}\ .
\end{equation}
Any vector $V^{\lambda}$ tangent to the hypersurface satisfies the relations
\be
V^{\lambda}n_{\lambda}=0\ , \quad V^{\lambda}=h^{\lambda}_{\;\;\sigma}V^{\sigma}\ .
\label{tangentvector}
\ee
whereas the 3-covariant derivative in the hypersurface is provided by the full projection of the space-time covariant derivative as
\be
^{(3)}\nabla_{\alpha}V^{\lambda}=h^{\beta}_{\alpha}h^{\lambda}_{\sigma}\nabla_{\beta}V^{\sigma}
\label{ProjDerivative}
\ee
The commutator of the 3-covariant derivatives acting on the tangent vector $V^{\lambda}$ provides the Riemann tensor of the hypersurface,
\be
\left[^{(3)}\nabla_{\alpha},^{(3)}\nabla_{\beta}\right]V^{\lambda}={}^{(3)}R^{\lambda}_{\; \; \delta\alpha\beta}V^{\delta}\ .
\label{Riemann3D}
\ee
Then, by using (\ref{ProjDerivative}) and after some algebraic manipulations, the commutator (\ref{Riemann3D}) can be expressed in terms of the Riemann tensor of the four-dimensional manifold as follows
\bea
&&\left[^{(3)}\nabla_{\alpha},^{(3)}\nabla_{\beta}\right]V^{\lambda}=\Big(h^{\alpha'}_{\; \alpha}h^{\beta'}_{\; \beta}h^{\lambda}_{\; \sigma}h^{\rho}_{\; \rho'}R^{\sigma}_{\; \; \rho\alpha'\beta'} \nonumber \\
&+&\epsilon\left(h^{\alpha'}_{\; \alpha}h^{\beta'}_{\; \beta}-h^{\alpha'}_{\; \beta}h^{\beta'}_{\; \alpha}\Big)h^{\lambda}_{\; \sigma}h^{\rho}_{\; \rho'}(\nabla_{\alpha'}n^{\sigma})(\nabla_{\beta'}n_{\rho})\right)V^{\rho'}\ .
\label{Commutator1}
\eea
In addition, as the covariant derivative of the normal vector is given by
\be
\nabla_{\mu}n^{\nu}=\partial_{\mu}n^{\nu}+L^{\nu}_{\;\; \mu\lambda}n^{\lambda}+A^{\nu}_{\;\; \mu\lambda}n^{\lambda}=\nabla^{L}_{\mu}n^{\nu}+A^{\nu}_{\;\; \mu\lambda}n^{\lambda}\ ,
\label{CovDeriv}
\ee
we might define the extrinsic curvature associated to the connection $L$ as
\be
K_{\mu\nu}(L)=h^{\alpha}_{\; \mu}h^{\beta}_{\; \nu}\nabla^{L}_{\alpha}n_{\beta}\ ,
\label{ExtrinsicL}
\ee
which can be straightforwardly shown to be a symmetric tensor. Finally, the full expression for the commutator (\ref{Commutator1}) leads to:
\begin{widetext}
\bea
\left[^{(3)}\nabla_{\alpha},^{(3)}\nabla_{\beta}\right]V^{\lambda}&=& h^{\alpha'}_{\; \alpha}h^{\beta'}_{\; \beta}h^{\lambda}_{\; \lambda'}h^{\rho}_{\; \rho'}R^{\lambda'}_{\; \; \rho\alpha'\beta'}V^{\rho'}
+\epsilon\left(K_{\; \; \alpha}^{\lambda}K_{\beta\rho'}-h^{\beta'}_{\; \beta}h^{\sigma}_{\; \rho'}A^{\delta}_{\;\; \beta'\sigma}K^{\lambda}_{\; \; \alpha}n_{\delta}+h^{\alpha'}_{\; \alpha}h^{\lambda}_{\; \lambda'}A^{\lambda'}_{\;\; \alpha'\delta}K_{\beta\rho'}n^{\delta}\right. \nonumber \\
&-&h^{\alpha'}_{\; \alpha}h^{\lambda}_{\; \lambda'}h^{\beta'}_{\; \beta}h^{\sigma}_{\; \rho'}A^{\lambda'}_{\;\; \alpha'\delta}A^{\delta'}_{\;\; \beta'\sigma}n_{\delta'}n^{\delta}-K_{\; \; \beta}^{\lambda}K_{\alpha\rho'}+h^{\beta'}_{\; \alpha}h^{\lambda'}_{\; \rho'}A^{\delta}_{\;\; \beta'\lambda'}K^{\lambda}_{\; \; \beta}n_{\delta}-h^{\alpha'}_{\; \beta}h^{\lambda}_{\; \lambda'}A^{\lambda'}_{\;\; \alpha'\delta}K_{\alpha\rho'}n^{\delta} \nonumber \\
 &+&\left. h^{\alpha'}_{\; \beta}h^{\lambda}_{\; \lambda'}h^{\beta'}_{\; \alpha}h^{\sigma}_{\; \rho'}A^{\lambda'}_{\;\; \alpha'\delta}A^{\delta'}_{\;\; \beta'\sigma}n_{\delta'}n^{\delta}\right) V^{\rho'}\ .
\label{Commutator2}
\eea
\end{widetext}
Hence, this expression allows us to write the Riemann tensor in the hypersurface in terms of the four-dimensional curvature tensor, the extrinsic curvature (\ref{ExtrinsicL}), and the 3-rank tensor $A^{\lambda}_{\;\; \mu\nu}$, leading to the expression
\begin{widetext}
\bea
{}^{(3)}R^{\lambda}_{\; \; \rho'\alpha\beta}&=& h^{\alpha'}_{\; \alpha}h^{\beta'}_{\; \beta}h^{\lambda}_{\; \lambda'}h^{\rho}_{\; \rho'}R^{\lambda'}_{\; \; \rho\alpha'\beta'}
+\epsilon\left(K_{\; \; \alpha}^{\lambda}K_{\beta\rho'}-h^{\beta'}_{\; \beta}h^{\sigma}_{\; \rho'}A^{\delta}_{\;\; \beta'\sigma}K^{\lambda}_{\; \; \alpha}n_{\delta}+h^{\alpha'}_{\; \alpha}h^{\lambda}_{\; \lambda'}A^{\lambda'}_{\;\; \alpha'\delta}K_{\beta\rho'}n^{\delta}\right. \nonumber \\
&-&h^{\alpha'}_{\; \alpha}h^{\lambda}_{\; \lambda'}h^{\beta'}_{\; \beta}h^{\sigma}_{\; \rho'}A^{\lambda'}_{\;\; \alpha'\delta}A^{\delta'}_{\;\; \beta'\sigma}n_{\delta'}n^{\delta}-K_{\; \; \beta}^{\lambda}K_{\alpha\rho'}+h^{\beta'}_{\; \alpha}h^{\lambda'}_{\; \rho'}A^{\delta}_{\;\; \beta'\lambda'}K^{\lambda}_{\; \; \beta}n_{\delta}-h^{\alpha'}_{\; \beta}h^{\lambda}_{\; \lambda'}A^{\lambda'}_{\;\; \alpha'\delta}K_{\alpha\rho'}n^{\delta} \nonumber \\
 &+&\left. h^{\alpha'}_{\; \beta}h^{\lambda}_{\; \lambda'}h^{\beta'}_{\; \alpha}h^{\sigma}_{\; \rho'}A^{\lambda'}_{\;\; \alpha'\delta}A^{\delta'}_{\;\; \beta'\sigma}n_{\delta'}n^{\delta}\right)\ .
\label{RiemannTensor3D4D}
\eea
\end{widetext}
In order to rewrite the action in terms of these objects, let us obtain the Ricci tensor of the independent connection and the corresponding curvature scalar. The former is given by
\bea
R_{\mu\nu} (\Gamma) &\equiv& R^{\lambda}_{\; \; \mu\lambda\nu}(\Gamma)=q^{\lambda\sigma}R_{\sigma\mu\lambda\nu}(\Gamma)\nn
&=&h^{\lambda\sigma}R_{\sigma\mu\lambda\nu}+\epsilon n^{\lambda}n^{\sigma}R_{\sigma\mu\lambda\nu}(\Gamma)\ .
\label{RicciTensor1}
\eea
while the latter reads as
\bea
R=q^{\alpha\beta}R_{\alpha\beta}(\Gamma)&=&h^{\alpha\beta}h^{\lambda\sigma}R_{\sigma\alpha\lambda\beta}(\Gamma)\nn
&+&\left[\epsilon\left(h^{\alpha\beta}n^{\sigma}n^{\lambda}+h^{\sigma\lambda}n^{\alpha}n^{\beta}\right)\right.\nn
&+&\left. n^{\alpha}n^{\beta}n^{\sigma}n^{\lambda}\right] R_{\sigma\alpha\lambda\beta}(\Gamma)\ .
\label{Riccicurvature}
\eea
The first term in the r.h.s of this expression is the projection of the Riemann tensor onto the hypersurface, which can be expressed in terms of the intrinsic and extrinsic curvature of the hypersurface through (\ref{RiemannTensor3D4D}) as follows:
\bea
&h^{\alpha\beta}&h^{\lambda\sigma}R_{\sigma\alpha\lambda\beta}(\Gamma)={}^{(3)}R+\epsilon\Big[K_{\alpha\beta}K^{\alpha\beta}-K^2 \nonumber \\
&-&\Big(K^{\alpha\beta}-h^{\alpha\beta}K\Big)A^{\delta}_{\;\; \alpha\beta}n_{\delta}+\Big(K^{\alpha}_{\;\; \lambda}-h^{\alpha}_{\;\; \lambda}K\Big)A^{\lambda}_{\;\; \alpha\delta}n^{\delta}\nonumber \\
&+&h^{[\alpha}_{\; \lambda}h^{\beta]\sigma}A^{\lambda}_{\;\; \alpha\delta}A^{\delta'}_{\;\; \beta\sigma}n^{\delta}n_{\delta'}\Big]\ ,
\eea
where $C^{[c}D^{d]}=C^{c}D^{d}-C^{d}D^{c}$. The last term in (\ref{Riccicurvature}) turns out zero by the symmetries of $R_{\sigma\alpha\lambda\beta}(\Gamma)$ while the second term can be obtained by using the split of the Riemann tensor (\ref{RiemannTensor}). The part depending uniquely on the connection $L$ reads:
\be
\epsilon\left(h^{\alpha\beta}n^{\sigma}n^{\lambda}+h^{\sigma\lambda}n^{\alpha}n^{\beta}\right)R_{\sigma\alpha\lambda\beta}(L)=2\epsilon n^{\alpha}n^{\beta}R_{\alpha\beta}(L)\ ,
\label{BoundL1}
\ee
where we have used some of the symmetry properties of the Riemann tensor. By using the identity $n^{\alpha}n^{\beta}R_{\alpha\beta}(L)=n^{\beta}\left[\nabla_{\alpha}^{L},\nabla_{\beta}^{L}\right]n^{\alpha}$, the expression (\ref{BoundL1}) leads to:
\bea
2\epsilon n^{\alpha}n^{\beta}R_{\alpha\beta}&=&2\epsilon \Big[\nabla_{\alpha}^{L}\Big(n^{\beta}\nabla_{\beta}^{L}n^{\alpha}-n^{\alpha}\nabla_{\beta}^{L}n^{\beta}\Big) \nonumber \\
&+&K^2-K_{\alpha\beta}K^{\alpha\beta}\Big]\ . \nonumber
\label{BoundL2}
\eea
Finally, the term involving the part of the Riemann tensor that depends on $A$ in (\ref{Riccicurvature}) yields,
\bea
\left(h^{\sigma\nu}n^{\lambda}n^{\mu}+h^{\lambda\mu}n^{\sigma}n^{\nu}\right)\mathcal{A}_{\lambda\sigma\mu\nu}(L)=\nn
=\nabla^{L}_{\mu}\Big[\Big(h^{\sigma[\nu}n^{\mu]}n_{\lambda}+h^{[\mu}_{\lambda}n^{\nu]}n^{\sigma}\Big)A^{\lambda}_{\;\,\sigma\nu}\Big]\nn
-A^{\lambda}_{\;\,\sigma\nu}\nabla^{L}_{\mu}\Big(h^{\sigma[\nu}n^{\mu]}n_{\lambda}+h^{[\mu}_{\lambda}n^{\nu]}n^{\sigma}\Big)\nn
+\Big(h^{\sigma\nu}n^{\mu}n_{\lambda}+h^{\mu}_{\lambda}n^{\nu}n_{\sigma}\Big)A^{\lambda}_{\;\,\rho[\mu}A^{\rho}_{\;\,\nu]\sigma}\ .
\label{BoundA1}
\eea
Finally, the full expression of the curvature scalar in terms of the extrinsic curvature and the projections onto the tangential and normal directions of the hypersurface is given by
\begin{widetext}
\bea
R&=&q^{\mu\nu}R_{\mu\nu}={}^{(3)}R+\epsilon\left\{-K_{\alpha\beta}K^{\alpha\beta}+K^2+\left(h^{\nu}_{\lambda}\nabla_{\mu}n^{\sigma}-h^{\nu\sigma}\nabla_{\mu}n_{\lambda}\right)n^{\mu}A^{\lambda}_{\sigma\nu}\right.\nonumber \\
&+&\left. h^{[\alpha}_{\; \lambda}h^{\beta]\sigma}A^{\lambda}_{\;\; \alpha\delta}A^{\delta'}_{\;\; \beta\sigma}n^{\delta}n_{\delta'}+\Big(h^{\sigma\nu}n^{\mu}n_{\lambda}+h^{\mu}_{\lambda}n^{\nu}n_{\sigma}\Big)A^{\lambda}_{\;\,\rho[\mu}A^{\rho}_{\;\,\nu]\sigma}\right.\nonumber \\
&+&\left. \nabla^{L}_{\mu}\Big[2\left(n^{\nu}\nabla^{L}_{\nu}n^{\mu}-n^{\mu}\nabla^{L}_{\nu}n^{\nu}\right)+\Big(h^{\sigma[\nu}n^{\mu]}n_{\lambda}+h^{[\mu}_{\lambda}n^{\nu]}n^{\sigma}\Big)A^{\lambda}_{\;\,\sigma\nu}\Big]\right\} \ .
\label{RicciScalar1}
\eea
\end{widetext}
The last term in this expression is a total derivative that under the action integral (\ref{eq:actionRBG}) yields a boundary term,
\begin{eqnarray}
\epsilon\int_{\mathcal{M}} d^4x \sqrt{-q}\nabla^{L}_{\mu}V^{\mu}=\int_{\Sigma} d^3y \sqrt{|h|} n_{\mu}V^{\mu}\ ,
\label{BSTheorem}
\end{eqnarray}
where
\be
V^{\mu}=2\left(n^{\nu}\nabla^{L}_{\nu}n^{\mu}-n^{\mu}\nabla^{L}_{\nu}n^{\nu}\right)+\Big(h^{\sigma[\nu}n^{\mu]}n_{\lambda}+h^{[\mu}_{\lambda}n^{\nu]}n^{\sigma}\Big)A^{\lambda}_{\;\,\sigma\nu}\ .
\label{TotalBoundary}
\ee
The first part of this boundary term is the usual GHY term (for the metric $q$), which after some manipulations gives
\be
2n_{\mu}\left(n^{\nu}\nabla^{L}_{\nu}n^{\mu}-n^{\mu}\nabla^{L}_{\nu}n^{\nu}\right)=-2\epsilon \nabla^{L}_{\nu}n^{\nu}=-2\epsilon K(L)\ ,
\label{GHY_term1}
\ee
whereas the second term in (\ref{TotalBoundary}) leads to
\bea
n_{\mu}\Big(h^{\sigma[\nu}n^{\mu]}n_{\lambda}+h^{[\mu}_{\lambda}n^{\nu]}n^{\sigma}\Big)A^{\lambda}_{\;\,\sigma\nu}=\nn
=\epsilon\left(q^{\sigma\nu}A^{\lambda}_{\;\,\sigma\nu}n^{\lambda}-n^{\sigma}A^{\lambda}_{\;\,\lambda\sigma}\right)\ .
\label{GHY_term2}
\eea
Therefore, the GHY boundary term that has to be added for a general RBG theory can be written as
\be
\mathcal{S}_{GHY}=\frac{1}{\kappa^2}\int_{\Sigma} d^3y \sqrt{|h|}\epsilon\left[K(L)-\frac{1}{2}\left(q^{\sigma\nu}A^{\lambda}_{\;\,\sigma\nu}n^{\lambda}-n^{\sigma}A^{\lambda}_{\;\,\lambda\sigma}\right)\right]\ .
\label{BoundaryFinal21}
\ee
Hence, in order to get a well formulated variational principle, this boundary term must be added to the gravitational action (\ref{eq:actionRBG}), leading to the so-called Gauss-Codazzi action for this type of theories
\begin{eqnarray}
\mathcal{S}_{RBG}&=& \frac{1}{2\kappa^2} \int d^4x \sqrt{-q} q^{\mu\nu}R_{\mu\nu}(\Gamma) \nonumber \\
&+&\frac{1}{\kappa^2}\int_{\Sigma} d^3y \sqrt{|h|}\epsilon\left[K(L)\right.\nn
&-&\left.\frac{1}{2}\left(q^{\sigma\nu}A^{\lambda}_{\;\,\sigma\nu}n_{\lambda}-n^{\sigma}A^{\lambda}_{\;\,\lambda\sigma}\right)\right]\ .
\label{Gauss-CodazziAction}
\end{eqnarray}
Note also that the field equations (\ref{eq:connsol}) yield $\Gamma^{\lambda}_{\;\;\mu\nu}=L^{\lambda}_{\;\;\mu\nu}$, i.e. the connection turns out to be the one compatible with the metric tensor $q$, such that $A^{\lambda}_{\;\;\mu\nu}=0$. It is important to notice also that one might have started by just assuming the connection to be compatible to a generic and arbitrary metric, which would have led to the same result. Indeed, by doing so, the gravitational action finally reads:
\begin{eqnarray}\label{Gauss-CodazziAction2}
\mathcal{S}_{RBG} &=& \frac{1}{2\kappa^2} \int d^4x \sqrt{-q} \Big[{}^{(3)}R+\epsilon\Big(-K_{\alpha\beta}K^{\alpha\beta}+K^2\Big) \nonumber \\ &+&2\nabla_{\alpha}\Big(n^{\beta}\nabla_{\beta}n^{\alpha}-n^{\alpha}\nabla^{L}_{\beta}n^{\beta}\Big)\Big] \\ 
 &+&2\int_{\Sigma} d^3y \sqrt{|h|}\epsilon K(L)\ .  \nonumber
\end{eqnarray}
This action is none other than the same action for GR with the GHY term but now defined in terms of the metric tensor $q$ instead of the space-time metric $g$. As shown in the next section, this might imply important and different consequences as compared to GR.

%%%%%%%%%%%%%%%%%%%%%%%%%%%%%%%%%%%%%%
\section{Covariant description of the boundary term revisited}
\label{CovariantSect}
%%%%%%%%%%%%%%%%%%%%%%%%%%%%%%%%%%%%%%

Let us now turn again to Eq.(\ref{eq:WrongST}) and reconsider it from the perspective of the splitting (\ref{eq:GammaSplit}), such that the variation $\delta \Gamma^\lambda_{\mu\nu}$ can be split as the sum of the variation of the fiducial connection $L^\lambda_{\mu\nu}$ plus the variation of the rank-3 tensor $A^\lambda_{\mu\nu}$, namely, $\delta \Gamma^\lambda_{\mu\nu}=\delta L^\lambda_{\mu\nu}+\delta A^\lambda_{\mu\nu}$. By construction, the contribution coming from the $q$-dependent $\delta L^\lambda_{\mu\nu}$ term is the usual one found in GR for a metric denoted $q_{\mu\nu}$, while the contributions due to the arbitrary connection become
\begin{eqnarray}\label{eq:surface1}
I_{\delta A}&= &\frac{1}{2\kappa^2}\int d^4x\sqrt{-q}\nabla^L_\alpha\left[q^{\mu\nu}\delta A^\alpha_{\mu\nu}-q^{\alpha\beta}\delta A^\lambda_{\lambda\beta}\right] \nonumber \\ &=& \frac{1}{2\kappa^2}\int d^3y\sqrt{h}\left[n_\alpha q^{\mu\nu}\delta A^\alpha_{\mu\nu}-n^\alpha\delta A^\lambda_{\lambda\alpha}\right] \ .
\end{eqnarray}
This is exactly the same expression with opposite sign that one obtains when varying the boundary term in (\ref{Gauss-CodazziAction}) with respect to the connection $A$. We are now interested in obtaining such boundary term associated to $\delta A^\lambda_{\mu\nu}$, necessary to compensate the $\delta A^\lambda_{\mu\nu}$ piece of Eq. (\ref{eq:surface1}) and its relation with the extrinsic curvature, by following \cite{BarberoG:2021cei}, where a general definition for the extrinsic curvature for an arbitrary connection $\Gamma$ is provided, 
\begin{equation}\label{eq:ExtrinsincK}
K_{\alpha\beta}(\Gamma)=\frac{1}{2}{h_\alpha}^\mu{h_\beta}^\nu\left[\nabla_\mu n_\nu+q_{\mu\lambda}\nabla _\nu n^\lambda\right] \ ,
\end{equation}
Then, this expression can be decomposed as 
\begin{equation}\label{eq:K_ab}
K_{\alpha\beta}(\Gamma)=K_{\alpha\beta}(L)-\frac{1}{2}{h_\alpha}^\mu{h_\beta}^\nu\left[A^\lambda_{\mu\nu} n_\lambda-q_{\mu\rho}A^\rho_{\nu\lambda}n^\lambda\right] \ ,
\end{equation}
and leads to 
\begin{equation}
K(\Gamma)= q^{\alpha\beta}K_{\alpha\beta}(\Gamma)=K(L)-\frac{1}{2}h^{\mu\nu}\left[A^\lambda_{\mu\nu} n_\lambda-q_{\mu\rho}A^\rho_{\nu\lambda}n^\lambda\right] \ .
\end{equation}
According to this, the extension of the GHY term to the Palatini case could be written as 
\begin{equation}\label{eq:GHY_Pal}
S_{GHY}=\frac{1}{\kappa^2}\int d^3x \sqrt{h}\left[K(L)-\frac{1}{2}h^{\mu\nu}\left[A^\lambda_{\mu\nu} n_\lambda-q_{\mu\rho}A^\rho_{\nu\lambda}n^\lambda\right]\right] \ .
\end{equation} 
Variation of this quantity with respect to $A^\rho_{\nu\lambda}$ leads to 
\begin{equation}
\delta_A S_{GHY}=-\frac{1}{2\kappa^2}\int d^3x \sqrt{h}\left[h^{\mu\nu}\delta A^\lambda_{\mu\nu} n_\lambda-h^{\mu\nu}q_{\mu\rho}n^\lambda\delta A^\rho_{\nu\lambda}\right] \ .
\end{equation}
Using in this expression the relation $h^{\mu\nu}q_{\mu\rho}=\delta^\nu_\rho -\epsilon n^\nu n_\rho$ and rewriting $h^{\mu\nu}=q^{\mu\nu}-\epsilon n^\mu n^\nu$, the result can be written as 
\begin{equation}
\delta_A S_{GHY}=-\frac{1}{2\kappa^2}\int d^3x \sqrt{h}\left[ n_\alpha q^{\mu\nu}\delta A^\alpha_{\mu\nu}-n^\alpha\delta A^\lambda_{\lambda\alpha}\right] \ ,
\end{equation}
which is exactly the term obtained in the previous section (\ref{Gauss-CodazziAction}) and compensates the variations given in (\ref{eq:surface1}). We thus see that, as shown in the previous section by other means, the generalization (\ref{eq:GHY_Pal}) of the GHY term does indeed compensate the boundary terms generated by the independent connection, both from the metric part and from the non-metric part. Note that the way this occurs is very interesting, as it implies that the $\delta A^\alpha_{\mu\nu}$ terms coming from the bulk action and from the GHY term cancel identically. In other words, unlike in Eq. (\ref{eq:WrongST}), one does not need to impose the vanishing of $\delta A^\alpha_{\mu\nu}$ on the boundary to have a well defined variational formulation because with the proposal (\ref{eq:GHY_Pal}) all such terms vanish identically for arbitrary variations.  Note also that the same results are obtained even in absence of any mapping of the gravitational action into a GR-like action as far as one splits the independent connection as (\ref{eq:GammaSplit}) and then, takes variations over the metric tensor $q$ and the independent 3-rank tensor $A$.

%%%%%%%%%%%%%%%%%%%%%%%%%%%%%%%%%%
\section{Hamiltonian formulation and ADM energy}
\label{Hamiltonian}
%%%%%%%%%%%%%%%%%%%%%%%%%%%%%%%%%%%

As in the GR case, the starting point for formulating the Hamiltonian approach is to introduce the ADM decomposition in order to foliate the space-time in a family of hypersurfaces of constant coordinate time $t$. The space-time metric for the ADM decomposition reads as \cite{Gourgoulhon:2007ue}:
\begin{eqnarray}
ds^2&=&g_{\mu\nu}dx^{\mu}dx^{nu} \nonumber \\
&=&-N^2dt^2+\gamma_{ij}\left(dx^{i}+N^{i}dt\right)\left(dx^{j}+N^{j}dt\right)\ .
\label{metricADM}
\end{eqnarray}
Here $N$ is the so-called lapse function, $N^{i}$ is the shift vector and $\gamma_{ij}$ is the induced metric on the hypersurfaces of constant time, given by
\be
\gamma_{\alpha\beta}=g_{\alpha\beta}+\tilde{n}_{\alpha}\tilde{n}_{\beta}\ ,
\label{ADM_ind_metric}
\ee
where $\tilde{n}^{\alpha}$ is the normal vector to the hypersurfaces. Nevertheless, for RBGs the gravitational action (\ref{Gauss-CodazziAction2}) does not depend directly on the space-time metric $g$ but on the metric tensor $q$. Both are related by (\ref{metricrselation}), which means that their relation can be expressed by the following transformation:
\be
q_{\mu\nu}= g_{\mu\alpha}\Omega^{\alpha}_{\;\;\nu}\ ,
\label{conform_transform}
\ee
where $\Omega^{\alpha}_{\;\;\nu}$ is a matrix that depends on the matter fields and possibly on $g_{\mu\nu}$ that can be obtained once the corresponding action in (\ref{metricrselation}) is provided. For instance, when considering the Eddington-inspired Born-Infeld (EiBI) theory of gravity, which is described by the Lagrangian density
\be
\mathcal{L}_{EiBI}=\frac{1}{\epsilon\kappa^2}\left(\sqrt{|g_{\mu\nu}+\epsilon R_{\mu\nu}|}-\lambda\sqrt{-g}\right)\ ,
\label{EiBILagrangian}
\ee
the expression (\ref{conform_transform}) can be worked out directly, leading to \cite{BeltranJimenez:2017doy}:
\be
g_{\mu\nu}=q_{\mu\nu}-\epsilon R_{\mu\nu}\ .
\ee
Moreover, by the EiBI field equations this relation can be expressed in a way that reflects better that the relation between metrics is modulated by the presence of matter distributions\footnote{The tilde on $\hat{T}_\mu\nu$ indicates that these are the matter fields in the Einstein frame of the theory. For details, see \cite{Olmo:2020fnk}.} \cite{Olmo:2020fnk}:
\be
g_{\mu\nu}=q_{\mu\nu}-\epsilon\kappa^2\left(\tilde{T}_{\mu\nu}-\frac{1}{2} g_{\mu\nu}\tilde{T}\right)\ .
\ee
 Note that, in general, the above relation between metrics is not of conformal type, but can have additional (disformal) terms. This type of generalized disformal relations is a generic property of RBGs, as can be seen in other models beyond the EiBI theory \cite{Magalhaes:2022esc,Afonso:2018hyj}. In order keep the discussion as general as possible, rather than focusing on specific gravity models, we will keep the relation between metrics as in (\ref{conform_transform}), so that the matrix $\Omega^{\alpha}_{\;\;\nu}$ is not given explicitly. For implementing the ADM decomposition, the metric tensor $q$ is expressed in a similar way as the spacetime metric tensor (\ref{metricADM}),
 \be
q_{\mu\nu}dx^{\mu}dx^{\nu}=-\mathcal{N}^2dt^2+h_{ij}\left(dx^{i}+\mathcal{N}^{i}dt\right)\left(dx^{j}+\mathcal{N}^{j}dt\right)\ .
\label{metricADMconfor}
\ee
From this metric, one finds directly that the spatial metric tensor $h$ is related with the metric tensor $q$ by:
\be
h_{\alpha\beta}=q_{\alpha\beta}+n_{\alpha}n_{\beta}\ .
\label{ADM_ind_metric_h}
\ee
Then, the relation between the induced spacetime metric (\ref{ADM_ind_metric}) and (\ref{ADM_ind_metric_h}) leads to (\ref{conform_transform}):
\be
\gamma_{\alpha\lambda}\Omega^{\lambda}_{\;\;\beta}-\tilde{n}_{\alpha}\tilde{n}_{\lambda}\Omega^{\lambda}_{\;\;\beta}= h_{\alpha\beta}-n_{\alpha}n_{\beta}\ .
\label{relationIndMetricsAndNormalVecs}
\ee
At this point, one might be tempted to relate both induced metrics by $h_{\alpha\beta}=\gamma_{\alpha\lambda}\Omega^{\lambda}_{\;\;\beta}$ (and a similar expression for the normal vectors) but this might not be the case in general, since the relations (\ref{conform_transform}) must be worked out for each specific case, such that the relation between both induced metrics is given by:
\be
h_{ij}=\gamma_{ik}\left(N^{k}\Omega^{0}_{\;\;j}+\Omega^{k}_{\;\;j}\right)\ .
\ee
On the other hand, the curvature scalar (\ref{RicciScalar1}) yields:
\be
R={}^{(3)}R+K^2+K_{ij}K^{ij}-\frac{2}{\mathcal{N}}\left(\partial_t K-\mathcal{N}^{i}\partial_{i}K\right)-\frac{2}{\mathcal{N}}\mathcal{D}_{i}\mathcal{D}^{i}\mathcal{N}\ ,
\label{Ricci_scalar_ADM}
\ee
where the covariant derivatives $\mathcal{D}_i={}^{(3)}\nabla_i$ are those along the hypersurface, compatible with the first fundamental form $h_{ij}$ and the terms depending on the connection tensor $A$ are omitted. On the other hand, the extrinsic curvature is\footnote{Here we have used the usual sign convention for the ADM decomposition, where the extrinsic curvature is defined as $K_{\mu\nu}(L)=-h^{\alpha}_{\; \mu}h^{\beta}_{\; \nu}\nabla^{L}_{\alpha}n_{\beta}$.}
\be
K_{ij}=-\frac{1}{\mathcal{N}}\left(\partial_{t}h_{ij}-\mathcal{D}_{i}\mathcal{N}_{j}-\mathcal{D}_{j}\mathcal{N}_{i}\right)\ ,
\label{extrinsicCurvatureADM}
\ee
Then, by using the ADM variables, the action (\ref{Gauss-CodazziAction2}), excluding the boundary contributions, becomes:
\begin{eqnarray}
S_{ADM}&=&\int dt \mathcal{L}_{ADM} \nonumber \\
&=& \int dt d^3x\sqrt{-h} \mathcal{N}\left[{}^{(3)}R-(K^2-K_{ab}K^{ab})\right]\ . 
\label{ADM_action1}
\end{eqnarray}
This is the ADM action for RBG theories, where $\mathcal{L}_{ADM}$ is the Lagrangian density. In order to construct the Hamiltonian, the corresponding momenta are obtained, where one gets the following constraints:
\bea
p_{\mathcal{N}}&=&\frac{\partial\mathcal{L}_{ADM}}{\partial{\dot{\mathcal{N}}}}=0 \\
p_{\mathcal{N}_a}&=&\frac{\partial\mathcal{L}_{ADM}}{\partial{\dot{\mathcal{N}_a}}}=0 \ .
\label{momentumConstraints}
\eea
While the conjugate momenta read as
\bea
\pi^{ab}&=&\frac{\partial\mathcal{L}_{ADM}}{\partial{\dot{h}_{ab}}}=\sqrt{h} (K_{ab}-h^{ab}K)  \\
\pi&=&h_{ab}\pi^{ab}=-2\sqrt{h}K \ ,
\label{conjugatemomentum}
\eea
Finally, by applying the Legendre transformation, the Hamiltonian density leads to:
\be
\mathcal{H}_{ADM}=\pi^{ab}\dot{h}_{ab}-\mathcal{L}_{ADM}=\mathcal{H}_{ADM}=\mathcal{N}\mathcal{H}+\mathcal{N}_a\mathcal{H}^a\ 
\label{ADM_Hamiltonian}
\ee
where
\be
\mathcal{H}=\sqrt{h}\left[Q^{abcd}K_{ab}K_{cd}-{}^{(3)}R\right]\ , \quad \mathcal{H}^{b}=-2\mathcal{D}_{a}\tilde{\pi}^{ab}\ .
\ee
By using the equations (\ref{momentumConstraints}), the variations over $\mathcal{N}$ and $\mathcal{N}^{a}$ become null:
\bea
\frac{\partial\mathcal{L}_{ADM}}{\partial\mathcal{N}}=\mathcal{H}=0\ ,\quad \frac{\partial\mathcal{L}_{ADM}}{\partial \mathcal{N}^{a}}=\mathcal{H}^{a}=0\ .
\label{ADM_constraints}
\eea
These are the Hamiltonian and momenta constraints in RBG theories, equivalent to those in GR. However, as all these magnitudes are related to the space-time ones by conformal transformations, this might have some consequences, as shown below.

Let us now obtain the ADM energy by following the standard procedure delineated in \cite{LectureNotes}. To do so, one has to consider the boundary contributions, since these are going to be the only non-null ones. Besides the ADM action (\ref{ADM_action1}), one has to account for the following boundary terms:
\be
S_{GHY}+S_{GC}=2\oint_{\mathcal{B}}d^2x \sqrt{h_{\mathcal{B}}}K_{B}+2\oint_{\mathcal{B}}d^2x\sqrt{h_{\mathcal{B}}}n^{\alpha}n^{\beta}\nabla_{\alpha}r_{\beta}\ ,
\label{BoundariesADM}
\ee
where $r_{a}$ is the normal vector to the boundary $\mathcal{B}$, while $h_{\mathcal{B}\mu\nu}=q_{\mu\nu}-r_{\mu}r_{\nu}$ is the induced metric and $K_{\mathcal{B}}$ the extrinsic curvature. Such boundary term arise due to the possible existence of time-like boundaries besides the space-like hypersurface defined by $n^{\mu}$. The first term in (\ref{BoundariesADM}) corresponds to the usual GHY boundary contribution, as obtained above. The second one comes from the Gauss-Codazzi decomposition (\ref{Gauss-CodazziAction2}), providing the contribution $r_{\alpha}\left(n^{\beta}\nabla_{\beta}n^{\alpha}-n^{\alpha}\nabla_{\beta}n^{\beta}\right)$, which after integrating by parts and using $r_\alpha n^{\alpha}=0$ leads to such a term that does not cancel with the GHY term. Both terms can be expressed as:
\be
S_{GHY}+S_{GC}=2\oint_{\mathcal{B}}d^2x \sqrt{h_{\mathcal{B}}} s^{\alpha\beta}\nabla_{\alpha}r_{\beta}
\ee
where $s^{\alpha\beta}=q^{\alpha\beta}+n^{\alpha}n^{\beta}-r^{\alpha}r^{\beta}$ and $h_{\mathcal{B}}=\mathcal{N}^2 \det[s]$. Then, the Hamiltonian with the boundary terms yields:
\bea
H_{ADM}^{Total}&=&H_{ADM}-\frac{1}{8\pi G}\oint d^2x\sqrt{s} \mathcal{N}K_s\nn
&+&\frac{1}{8\pi G}\oint d^2x \mathcal{N}_{a}\pi^{ab}r_b\ .
\eea
Here $H_{ADM}=\int d^3x\ \mathcal{H}_{ADM}=\int d^3x\ \left(\mathcal{N}\mathcal{H}+N_a\mathcal{H}^a\right)$ and $K_s=s^{\alpha\beta}\nabla_{\alpha}r_{\beta}$. By applying the Hamiltonian constraints (\ref{ADM_constraints}), one finally arrives to the final result for the Hamiltonian density as
\be
H_{ADM}^{Total}=-\frac{1}{8\pi G}\oint d^2x\sqrt{s} \mathcal{N}K_s+\frac{1}{8\pi G}\oint d^2x \mathcal{N}_{a}\pi^{ab}\tilde{r}_b\ .
\label{HADM4}
\ee
For an asymptotic observer, $N\rightarrow 1$ and $N^{i}\rightarrow 0$ in the spacetime metric (\ref{metricADM}), while the values for $\mathcal{N}$ and $\mathcal{N}^{i}$ asymptotically are obtained through (\ref{conform_transform}), leading to:
\bea
-\mathcal{N}^2+h_{ij}\mathcal{N}^{i}\mathcal{N}^{j}&=&-\Omega^{0}_{\;\;0}\ , \quad
2h_{ij}\mathcal{N}^{j}=-\Omega^{0}_{\;\;i}\ , \nn
h_{ij}&=&\gamma_{ik}\Omega^{k}_{\;\;j}\ .
\label{relationsMetricsAsymp}
\eea
Note that in comparison to GR, the second integral in (\ref{HADM4}) might not be removed when taking the asymptotic limit $N^{i}\rightarrow 0$ unless $\Omega^{0}_{\;\;i}=0$, such that the second integral might contribute to the ADM energy. In any case, in order to get a well defined ADM energy, the contribution at spatial infinity must be subtracted in both integrals in (\ref{HADM4}). For the more familiar case where $\mathcal{N}^{j}\rightarrow 0$ asymptotically, then the expression for the ADM energy is obtained by subtracting the contribution $K_s^{0}$ at spatial infinity, such that the ADM energy reads:
\be
E_{ADM}=-\frac{1}{8\pi G}\lim_{R\rightarrow\infty}\oint d^2x\sqrt{s}\Omega^{0}_{\;\;0}\left(K_s-K_s^{0}\right)\ ,
\label{E_ADM}
\ee
However, even in this case and despite that the expression for ADM energy looks as in GR, they do not lead to the same results. Firstly, due to the presence of the factor $\Omega^{0}_{\;\;0}$, and secondly, because both the extrinsic curvature and the induced metric given in the integral in (\ref{E_ADM}) do not describe an slice of the space-time but an hypersurface of the manifold defined in terms of the metric tensor $q_{\mu\nu}$, such that the ADM energy might not coincide with the energy associated to the matter fields of a particular system but it also contains contributions from the gravitational sector. 

To illustrate the last point let us apply this procedure to a simple case: the ADM energy for the Schwarzschild black hole in RBG theories. In the usual asymptotic spherical coordinates ($t,r,\theta,\varphi$), the line element for Schwarzschild space-time is described by the metric $g_{\mu\nu}$ given by (in units of $c=1$):
\begin{eqnarray}
ds^2&=&g_{\mu\nu}dx^{\mu}dx^{\nu}\\
&=&-\left(1-\frac{2GM}{r}\right)dt^2+\left(1-\frac{2GM}{r}\right)^{-1}dr^2+r^2d\Sigma^2\ .  \nonumber 
\label{Schw-metric}
\end{eqnarray}
where $d\Sigma=d\theta^2+\sin^2\theta d\varphi^2$ is the line element of the 2-sphere. This space-time metric might be a solution of the corresponding RBG theory as far as $R_{\mu\nu}(q)=0$ holds in the field equations. For this simple case, in absence of matter fields the matrix $\Omega^{\mu}_{\;\;\nu}$ in (\ref{conform_transform}) reduces to a conformal factor $\Omega^{\mu}_{\;\;\nu}=\Omega^2 \delta^{\mu}_{\;\;\nu}$. Such a conformal factor $\Omega^2$ depends upon derivatives of the action which at the same time depends on contractions and/or products of the Ricci tensor, which is null in this case, and consequently the conformal factor becomes a constant. In order to calculate the ADM energy given in (\ref{E_ADM}), the corresponding extrinsic curvature defined by the metric tensor $q$ must be obtained and also the corresponding first fundamental forms for the asymptotic hypersurface for both the conformal related metric of (\ref{Schw-metric}) as for the conformal Minkowski. The latter is the contribution that must be subtracted in (\ref{E_ADM}), since the metric (\ref{Schw-metric}) is asymptotically flat, which is given in same coordinates as (\ref{Schw-metric}) by: 
\be
ds^2=\eta_{\mu\nu}dx^{\mu}dx^{\nu}=-dt^2+dr^2+r^2d\Sigma^2\ .
\label{Schw-metric}
\ee
The corresponding induced metric tensor in the hypersurface of constant time is given by:
\begin{eqnarray}
\gamma_{ij} &\simeq & \left(1+\frac{2GM}{r}\right)dr^2+r^2d\Sigma^2 \nonumber \\
& \rightarrow & \quad h_{ij}\simeq \Omega^2\left(1+\frac{2GM}{r}\right)dr^2+(\Omega r)^2d\Sigma^2\ .
\label{indmetrics}
\end{eqnarray}
On the other hand, the normal vector $r^{\alpha}$ to the boundary $r=R$ is:
\be
r_{\alpha}\simeq\Omega\left(1+\frac{GM}{r}\right)\delta^{r}_{\alpha}\ .
\ee
Finally, the induced metric to such a boundary is just the metric of the 2-sphere $s_{AB}dy^{A}dy^{B}=(\Omega r)^2d\Sigma^2$ both for the Schwarzschild case and for the asymptotic Minkowski space-time. The extrinsic curvature for both can be easily obtained as:
\be
K_s=\frac{2}{R}\left(1-\frac{GM}{R}\right)\frac{1}{\Omega}\ , \quad K_s^0=\frac{2}{R}\frac{1}{\Omega}\ ,
\label{extrinsicShwMink}
\ee
Finally, the integral of the ADM energy (\ref{E_ADM}) yields
\begin{eqnarray}
E_{ADM}&=&-\frac{1}{8\pi G}\lim_{R\rightarrow\infty}\oint\Omega^2 R^2\sin\theta d\theta d\varphi  \left(-\frac{2GM}{R^2}\frac{1}{\Omega}\right) \nonumber \\
&=&\Omega M\ .
\label{E_ADM}
\end{eqnarray}
As shown above, the ADM energy now has a contribution on the conformal factor $\Omega$ that depends on the gravitational action. This fact might lead to differences in the ADM energy as far as such a factor is different from unity in specific solutions of RBGs as compared to their GR counterparts.

%%%%%%%%%%%%%%%%%%%%%%%%%%%%%%%%%%%%%%
\section{Conclusions}
\label{conclusions}
%%%%%%%%%%%%%%%%%%%%%%%%%%%%%%%%%%%%%%

In this paper, the Hamiltonian formulation of RBG theories has been obtained together with an analysis of its boundary terms. To do so, we have started by using the corresponding mapping proposed in \cite{Afonso:2018bpv}, by which the space-time metric is replaced in the action by another metric tensor that leads to a linear Lagrangian with respect to the Ricci tensor and a similar form as the usual Hilbert-Einstein action coupling to a modified matter sector. Such a tool has turned out to very useful as it allows to both analyse formal aspects and find explicit solutions of these theories in a much simpler way. 

The decomposition of the connection $\Gamma_{\mu\nu}^{\lambda}$ given in Eq. (\ref{eq:GammaSplit}) allowed us to interpret its variation as the variation of the fiducial connection $L_{\mu\nu}^{\lambda}$ plus the variation of the rank-3 tensor $A_{\mu\nu}^{\lambda}$. This puts forward that the variation of a connection cannot be regarded as just the variation of an ordinary tensorial field, because in that case one would miss the additional contribution that we found in $L_{\mu\nu}^{\lambda}$, which is crucial to obtain the correct boundary terms. This clarifies the apparent inconsistency pointed out at the end of section II, when the variation was implemented assuming that $\delta \Gamma_{\mu\nu}^{\lambda}\sim \delta A_{\mu\nu}^{\lambda}$.

Anchored on this procedure, we built the Hamiltonian formulation first by expressing the Riemann tensor in terms of projections onto a (family of) hypersurface(s), leading to the $3+1$ decomposition of the theory by expressing the Ricci tensor/scalar in terms of the extrinsic and intrinsic curvature of the hypersurface. As a consequence, some new terms might arise in comparison to those of GR, as the presence of a rank-3 tensor in the connection components introduces new terms in such decomposition. However, since the field equations make such tensor null, similar expressions as those of GR are obtained. However, one should note that such decomposition is now performed in terms of a metric tensor that does not coincide with the space-time metric, but are instead related by a tensorial transformation. The corresponding GHY boundary term is obtained for any RBG theory, which in particular recovers the one obtained previously in the literature for the particular $f(R)$ gravity case \cite{Saez-ChillonGomez:2020afj}.

Next, by following the standard procedure as in GR, the ADM decomposition as well as the Hamiltonian formulation of the theory were carried out. The Hamiltonian constraints are then obtained, which turn out to coincide with the ones in GR, making the boundary terms the only non-zero contributions to the ADM Hamiltonian. The point is that for the general case, the ADM energy is sourced by the two boundary integrals as given in (\ref{HADM4}), contrary to the GR case where the second integral cancelled out. Such contributions now depend on the extrinsic curvature as defined in terms of the connection compatible to the metric tensor $q_{\mu\nu}$ and, consequently, it includes contributions that depend on the derivatives of the action through the (matter-dependent) matrix that relates both metrics via Eq.(\ref{metricrselation}). While such a result was expected, as it already occurs for instance in metric and Palatini $f(R)$ theories with the boundary contributions, it also affects the ADM energy. Indeed, as shown for the simple case of Schwarzschild space-time where the contribution from the second integral is removed, this leads to a contribution on the ADM energy that would vary from one action to other. In addition, the way this might be interpreted can be as follows: the contribution to an effective gravitational constant $G_{eff}=\Omega G$, such that the energy is then given by the mass of the central object. However, this effective gravitational constant will vary from one system to another. On the other hand, the ADM energy (\ref{E_ADM}) might be read off as a sort of self-contribution of gravity to the energy. The latter would lead to some exotic consequences as possible naked singularities keeping the energy of the system positive.

To conclude, this paper represents the closure of a number of technical issues that remained to be settled in the formulation of RBGs, and further contributes to the theoretical and observational viability of such theories. Indeed, these results are added to the pool of appealing properties of RBGs already found throughout the literature, such as the second-order field equations, absence of ghosts, natural recovery of the weak-field limit to make them compatible with GR tests, or the propagation of two tensorial polarizations of the gravitational field (gravitational waves) at the speed of light. Some open questions remained to be dealt with, though, as the extension of the results to other more complex solutions rather than the simple Schwarzschild space-time.

%%%%%%%%%%%%%%%%%%%%%%%%%%%%%%%%%%%
\section*{Acknowledgements}
%%%%%%%%%%%%%%%%%%%%%%%%%%%%%%%%%%%

This work is supported by the Spanish National Grants FIS2017-84440-C2-1-P, PID2020-116567GB-C21, PID2020-117301GA-I00, and PID2022-138607NB-I00, funded by MCIN/AEI/10.13039/501100011033 (``ERDF A way of making Europe" and ``PGC Generaci\'on de Conocimiento"), and the project PROMETEO/2020/079 (Generalitat Valenciana).

\end{document}